# Coupling light and sound: Giant nonlinearities from oscillating bubbles and droplets

Ivan S. Maksymov[1,2,3] and Andrew D. Greentree[1]
[1]Australian Research Council Centre of Excellence for Nanoscale BioPhotonics, School of Science, RMIT University, Melbourne, Victoria 3001, Australia
[2]Centre for Micro-Photonics, Swinburne University of Technology, Hawthorn, VIC 3122, Australia
[3]imaksymov@swin.edu.au

**Abstract**: Nonlinear optical processes are vital for fields including telecommunications, signal processing, data storage, spectroscopy, sensing, and imaging. As an independent research area, nonlinear optics began with the invention of the laser, because practical sources of intense light needed to generate optical nonlinearities were not previously available. However the high power requirements of many nonlinear optical systems limit their use, especially in portable or medical applications, and so there is a push to develop new materials and resonant structures capable of producing nonlinear optical phenomena with low-power light emitted by inexpensive and compact sources. Acoustic nonlinearities, especially giant acoustic nonlinear phenomena in gas bubbles and liquid droplets, are much stronger than their optical counterparts. Here, we suggest employing acoustic nonlinearities to generate new optical frequencies, thereby effectively reproducing nonlinear optical processes without the need for laser light. We critically survey the current literature dedicated to the interaction of light with nonlinear acoustic waves and highly-nonlinear oscillations of gas bubbles and liquid droplets. We show that the conversion of acoustic nonlinearities into optical signals is possible with low-cost incoherent light sources such as light-emitting diodes, which would usher new classes of low-power photonic devices that are more affordable for remote communities and developing nations, or where there are demanding requirements on size, weight and power.

## 1. Introduction

Maxwell's equations define and shape our lives: describing the physics from the electromagnetic radiation from the Sun to the interactions of atoms [1]. They are also essential for understanding and controlling light, thereby leading to the advanced photonic devices that are so critical to our lives and global economy: optical fibres, resonators, switches, diffraction gratings, displays, and many others [2-4].

Maxwell's equations are intrinsically linear, and in most cases we may assume that all optical interactions are also linear. We can say that when red light passes through an optical medium, for example a pair of glasses, it will still be red when it exits the medium. The linearity of optical processes allows us to describe the physics of light separately from the physics of the medium. The medium modifies light, but light does not change the medium.

The possibility to produce nonlinear optical effects with intense light was raised in the 1920-40s (see [5, 6]). However, nonlinear optics was established only with the invention of the laser [4, 5, 7, 8]. When the intensity of light is sufficiently strong, light can change the properties of the medium it is travelling through, which in turn alters the optical field. In this regime, it is no longer possible to treat the light and medium separately. These mutual light-matter interactions lead to new optical effects [8] and new nonlinear photonic devices [9-12].

One of the most important nonlinear processes is second harmonic generation [5, 7, 8], where high-intensity pump light enters a nonlinear optical material and a weak optical beam is generated at a frequency twice that of the original pump beam. Other important nonlinear processes are third harmonic generation, sum- and difference-frequency generation, and intensity-dependent index of refraction (Kerr effect) to name a few [8].

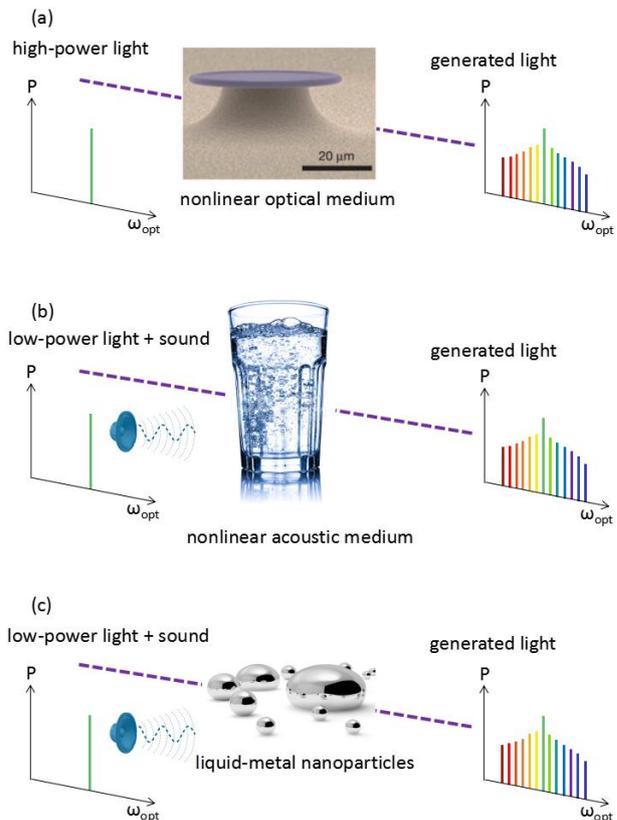

*FIG. 1*: *(a) Generation of new optical frequencies via a nonlinear optical interaction. High-power monochromatic laser light interacts with the nonlinear optical medium, e.g. a dielectric microresonator, to generate new optical frequencies [9]. (b, c) The nonlinear properties of sound can be used to generate optical nonlinearities without the need for high-power laser light. Low-power light couples to a sound wave via a deformable fluid, e.g. gas bubble or liquid-metal nanoparticle, exploiting strong nonlinear acoustic effects. This converts acoustic frequencies to the optical domain, effectively reproducing the result of the nonlinear optical interaction in (a). Images from www.freeimages.com and www.dreamstime.com.*

High-power lasers required to produce nonlinear effects are, however, expensive, energy-consuming, and the high intensity of light produced by them can cause damage to live cells and tissue, raising significant health and safety issues [13, 14]. This has motivated the research community to seek alternative solutions for the generation of optical nonlinearities with low-power light.

There have also been mostly sporadic attempts to achieve nonlinear optical phenomena with low power, incoherent light or at least to reproduce functionality of some nonlinear photonic devices without lasers [15-25]. However, even though the number of such works is gradually increasing and they attract keen interest of researchers in the field of nonlinear optics, they have not had a significant impact yet. This is because the fulfillment of the conditions of phase matching and high optical intensity with incoherent light is even more challenging than with laser light.

This does not imply though that phase matching cannot be satisfied with incoherent light. For example, pioneering experiments demonstrated sum-frequency generation via mixing of a ruby laser signal with an incoherent signal produced by a mercury lamp [15]. In that paper, exactly the same technique as in contemporary 'all-laser' experiments (special orientation of a KDP crystal [2]) was employed to satisfy phase matching. However, experimental schemes using non-laser light were mostly abandoned in the late 1960s because lasers operating at different wavelengths became available.

The requirement for high-intensity light is also difficult to satisfy with incoherent light. Consider a 1 W pulsed laser and a 1 W light-emitting diode (or a gas discharge lamp). One can generate optical nonlinearities with the laser because light emitted by it is coherent and can therefore be focused so that the medium experiences very high instantaneous intensities. Even though the light-emitting diode (gas discharge lamp) has the same time averaged power, the maximum intensity of light focused using the same optical components is fundamentally limited [26] by laws of thermodynamics and the concept of optical étendue defining how 'spread out' the light is in area and angle [2].

In this paper, we introduce a conceptually different approach — photonic devices capable of generating new optical frequencies with low-power light and nonlinear processes of non-optical origin. Here we suggest to use low-power, incoherent light to harness giant nonlinear acoustic properties of gas bubbles and liquid droplets, and exploit them in the optical domain. Effectively, this will reproduce the behaviour that is usually achieved via a nonlinear optical interaction enabled by high-power laser light.

Sound is described by travelling pressure waves that result from the motion of particles in the medium through which the sound wave is propagating. Although there are fundamental differences between optics and acoustics, it is possible to treat sound propagation using methods analogous to those used for light. In particular, concepts like wavelength, speed of propagation, and intensity, all have direct analogues between optics and acoustics. But while light fields typically do not produce large changes in the optical medium, with sound, atoms are being moved relatively large distances even at low intensities, creating non-uniform pressure fields and hence non-uniform speeds of sound in the medium. This in turn means that acoustic nonlinearities are far easier to observe than optical nonlinearities at comparable power levels [27, 28].

As a practical example, we first consider the optical frequency comb [Fig. 1(a)]. An optical frequency comb is a spectrum consisting of a series of discrete, equally spaced elements that have a well-defined phase relationship between each other [29-32]. Frequency combs have found important practical applications in precision measurements, microwave generation, telecommunications, astronomy, spectroscopy and imaging [31-34]. Combs have typically been produced by mode-locked lasers or exploiting nonlinear optical effects in optical fibres [29-32] and nonlinear photonic microresonators [9, 11, 12, 35] (see Fig. 1(a) and Section 2).

Acoustic nonlinearities provide an alternate low-power route to generating an optical frequency comb [Fig. 1(b, c)], which we discuss more completely in Section 3.4. This effectively reproduces the result of a classical nonlinear optical interaction [Fig. 1(a)]. High power light is not required for the conversion of acoustic nonlinearities into optical signals. This implies that our approach would, in general, allow the use of low-cost, low-power incoherent optical sources such as light-emitting diodes and discharge lamps for tasks such as optical frequency comb metrology.

This paper is organised as follows. In Section 2, we overview the recent advances in the research direction of low-power nonlinear optics. Section 3 introduces the concept of reproduction of nonlinear optical effects with nonlinear sound and low-power light. The discussion in Section 3 is supported by theory and numerical simulations demonstrating the possibility to generate optical frequency combs from nonlinear ultrasound. In Section 4, we discuss the giant acoustic nonlinearities of liquid droplets and gas bubbles in liquids. We also overview experimental techniques enabling the conversion of giant acoustic nonlinearities into the optical domain. Section 5 focuses on the possibility to employ plasmonic properties of optical nanoantennas and liquid-metal nanoparticles to enhance the interaction between nonlinear sound and light. Finally, in Section 6, we explore novel opportunities enabled by the interaction of light with extreme nonlinear behaviour of acoustically-driven gas bubbles.

2. **Low-power nonlinear photonics**

To achieve nonlinear optical processes with low-power laser light, one can (i) employ nonlinear optical materials with an intrinsically high nonlinear response and/or (ii) create resonant structures to locally increase the light-matter interactions that lead to stronger nonlinear effects. Strong nonlinear optical effects can be achieved in non-resonant photonic structures such as dielectric slot waveguides.

Despite the range of strategies available, in practice, the choice of nonlinear optical architectures is often limited by peculiarities of the available fabrication techniques for the desired optical material, toxicity (important for biomedical applications), cost and other factors. Here, one may employ materials with high optical nonlinearities such as chalcogenide glasses [36, 37], silicon [38, 39], and AlGaAs [40, 41]. Alternatively, one can exploit optical nonlinearities of noble metals such as gold [42, 43, 44]. However, factors such as immature fabrication technology for chalcogenide glass, high linear and nonlinear losses for semiconductors, and high losses in metal structures due to the excitation of plasmon resonances [45] motivate the search for alternative materials such as silica-based glasses [46] and indium tin oxide [47].

A large and growing body of research investigates high-quality factor (high-Q) nonlinear photonic micro-resonators [9, 48-63], integrated optical waveguides [64-67], optical nanoantennas [42, 68-80], and metasurfaces [81-84]. A judicious design of such structures enables large local enhancements of the optical field intensity, which in turn leads to strong nonlinear optical effects.

However, there also are significant drawbacks with the use of high-Q micro-resonators. Firstly, high intensity of light trapped inside a high-Q resonator can give rise to considerable multiphoton absorption losses. Indeed such losses are a significant obstacle for silicon-based photonic devices operating at telecommunication and mid-infrared wavelengths [85, 86]. Secondly, although low-power nonlinear effects are possible mostly with ultra-high-Q microresonators, such devices are sensitive to even minor fabrication imperfections, which compromises their scalability and on-chip integrability. High-Q microresonators excited at their resonance frequency are also sensitive to external perturbations, which significantly complicates their operation in real-life conditions.

Consequently, alternative strategies to achieve optical nonlinearities with low-power light have been proposed. For example, it was suggested that strong nonlinear optical effects suitable for optical comb generation may be achieved with low-power laser light in micro-ring resonators operating in a period-doubling bifurcation regime [87]. In Ref. [88], it was suggested that a single quantum dot integrated with a micropillar resonator [89] can be used to generate optical frequency combs with ultra-low power laser light of just of few nW. However, these new ideas still employ regular nonlinear optical processes and materials and require a source of coherent light. Consequently, they also inherit some of the fundamental limitations of modern nonlinear photonic devices such as high optical losses, sensitivity to fabrication imperfection, toxicity of nonlinear optical materials, and so on.

## 3. Photonics with nonlinear sound

It is instructive to start our discussion with a brief overview of the physics that underpins the nonlinear processes in optics and acoustics. We will use similar mathematical approaches to analyse optical and acoustic waves in their respective nonlinear media. This will open up an opportunity to compare the strength of the nonlinear effects.

### 3.1. Origin of nonlinear optical effects

By considering Maxwell's equations and taking into account the polarisation term, written for simplicity in scalar form as $P = \varepsilon_0(\chi_0 + \chi^{(1)}E + \chi^{(2)}E^2 + \chi^{(3)}E^3 + …)$, where the $\chi^{(j)}$ are the nonlinear susceptibilities, it can be shown that the polarisation of the material responds with powers of the electric field $E$. If we assume that $E = E_0\exp(i\omega t)$, where $\omega$ is the frequency of the incident light and $t$ is time, then the polarisation is $P(t) = \varepsilon_0\chi_0 + \varepsilon_0(\chi^{(1)}E_0 e^{i\omega t} + \chi^{(2)}E_0^2 e^{2i\omega t} + \chi^{(3)}E_0^3 e^{3i\omega t} + …)$. By grouping all the terms in powers of two and higher into the nonlinear polarisation $P_{NL}$, we can arrive to the nonlinear wave equation $\nabla^2 E - (n^2/c^2)\partial^2 E/\partial t^2 = -T$, where $n$ is the optical refractive index, $c$ is the speed of light, and $T = -\mu_0 \partial^2 P_{NL}/\partial t^2$ is the source term.

So the system responds as an anharmonic oscillator and the polarisation term contains not just the fundamental frequency $\omega$, but also multiples of this frequency (and in general a dc component). For example, in the original second harmonic generation experiment [7], due to the anharmonic response of the medium the high-intensity red light produced by a ruby laser (~694 nm wavelength) generates ultraviolet light (~347 nm), which is twice the frequency of the incident light.

Furthermore, if we consider a bichromatic optical field (with the angular frequencies $\omega_1 > \omega_2$) incident on a $\chi^{(2)}$ material, we can demonstrate that the source term $T$ of the nonlinear wave equation has four components, namely $2\omega_1$, $2\omega_2$, $(\omega_1 + \omega_2)$, and $(\omega_1 - \omega_2)$, which give rise to sum- and difference-frequency generation. More generally, one can demonstrate the ability to generate extra terms so that the process can be seen as a natural consequence of the nonlinear interactions. These complex processes may be used, for example, to generate optical frequency combs [11, 12, 79].

### 3.2. Origin of nonlinear acoustic effects

The origin of acoustic nonlinearities is qualitatively similar, but different in terms of the magnitude of the effect. For a sound wave travelling through air or water, the vibrations of the particles of the medium are best described by longitudinal waves [27]. (Optical waves are transverse, but this distinction does not affect the generality of our discussion because in solids sound can be transmitted as both longitudinal waves and transverse waves, also giving rise to strong nonlinear effects [28].)

By analogy with the nonlinear wave equation for optical waves, we can write a nonlinear wave equation for acoustic waves by considering the system of Euler equations [27]. We can arrive to the nonlinear wave equation $\partial^2 v/\partial t^2 - c_0^2 \partial^2 v/\partial x^2 = -T_a$, where $v$ is the particle velocity, $c_0$ is the ambient speed of sound, and the source term $T_a = L_2(v^2) + L_3(v^3) + L_4(v^4) + \ldots$ Here, $L_2(v^2)$ contains the terms in powers of two, $L_3(v^3)$ the terms in powers of three, and so on. For example, in explicit form $L_2(v^2) = v^2 + 0.5(\gamma-1)c_0^{-2}\left(\int[\partial v/\partial t]dx\right)^2$ [90], where $\gamma = c_p/c_v$ being $c_v$ and $c_p$ the specific heat capacities at constant volume and pressure, respectively. In the following we will demonstrate that $\gamma$ defines the strength of the nonlinear response.

We find solutions for a sinusoidal plane wave $v = v_0\sin(\omega_a t)$ originating from the point $x = 0$ and propagating along the $x$-coordinate direction. Here, $\omega_a$ is the acoustic angular driving frequency and $t$ is time. By expressing $v = v^{(1)} + v^{(2)} + v^{(3)} + \ldots$, in the second approximation we can write $\partial^2 v^{(2)}/\partial t^2 - c_0^2 \partial^2 v^{(2)}/\partial x^2 = L_2 v_0^2 \sin^2\omega_a t' + L_3 v_0^3 \sin^3\omega_a t' + \ldots$, where $t' = t - x/c_0$. Thus, as in the case of optical waves, the nonlinear acoustic medium acts as an anharmonic oscillator and its response contains not just the fundamental frequency $\omega_a$, but also multiples of this frequency (and also a dc acoustic component).

Similar to the case of optical waves, considering only the quadratic acoustic nonlinearity term and a bi-harmonic acoustic wave with the angular frequencies $\omega_{1(a)} > \omega_{2(a)}$, we can demonstrate that the source term $T_a$ of the nonlinear acoustic wave equation has four components, namely $2\omega_{1(a)}$, $2\omega_{2(a)}$, $(\omega_{1(a)} + \omega_{2(a)})$, and $(\omega_{1(a)} - \omega_{2(a)})$, which gives rise to sum- and difference-acoustic frequency generation. More generally, one can demonstrate the ability to generate extra terms so that the process can be seen as a natural consequence of the nonlinear acoustic interactions.

Let us now assume that, without loss of generality, the plane acoustic wave travels in the positive $x$-direction. Using the equations derived above, one can show [27] that the propagation speed for any particular point of this wave is given by the local value of $(c_0 + \beta v)$, where $\beta = (\gamma+1)/2$ is the nonlinear acoustic parameter. Hence, the points with $v > 0$ (e.g. wave crests) travel faster than $c_0$ and correspond to areas of compression. (Recall that longitudinal acoustic waves can be described as alternating areas of compression and rarefaction.) Conversely, the points with $v < c_0$ (e.g. wave troughs) travel slower and correspond to areas of rarefaction. This is opposed to the case of an idealised, linear wave propagation with $\beta = 0$, where all points of the wave travel at $c_0$ [compare the solid and dashed curves in Fig. 2(a)].

Thus, an initially sinusoidal sound wave undergoes deformations as it propagates, which implies that its initial monochromatic spectrum [Fig. 2(b)] gets enriched with new higher harmonic frequencies [91, 92] [Fig. 2(c)]. This spectral enrichment is stronger in acoustic media with larger values of the nonlinear acoustic coefficient $\beta$, the physical meaning of which is as follows. The rate of deformation of the propagating acoustic wave depends on the distance between its crests and troughs, and this depends on both the initial particle velocity $v_0$ and the value of $\beta$ for the particular fluid. For example, water has $\beta = 3.5$ and air has $\beta \approx 0.7$ [27, 28]. As a result of this dependence, a fluid with a higher $\beta$ will exhibit more rapid wave deformation than a fluid with a lower $\beta$, provided the same initial velocity $v_0$ is applied to both.

Significantly, an acoustic medium with $\beta = 0$ is just an idealisation. Hence, even in the low-power acoustic regime characterised by small values of $v_0 \ll c_0$, there is no ideal linear propagation and, as with the case of nonlinear optics, the acoustic medium responds anharmonically.

Despite isolated attempts to directly compare the strength of nonlinear optical and nonlinear acoustic effects [93], this task is, in general, challenging due to the different nature of light and sound. Indeed, one could compare the peak amplitude of the electric field $E_0$ with the peak particle velocity $v_0$. However, nonlinear optical effects start to manifest themselves only when optical electric field amplitudes reach high values of $10^8 \ldots 10^{10}$ V/m [2]. This is in stark contrast with nonlinear acoustics where systematic observation of practicable nonlinear acoustic effects is always possible, even with a very small $v_0$.

Alternatively, one could also compare both types of nonlinearity by comparing the mathematical role of the nonlinear acoustic parameter $\beta$ and the nonlinear optical susceptibilities $\chi$ in their respective wave equations. While successful in some particular cases (e.g. when optical nonlinearities of dielectrics are compared to acoustic nonlinearities of bulk fluids [93]), this approach fails in other important situations. The strength of nonlinear acoustic processes also depends on the compressibility of the medium, which is a measure of the relative volume change of the medium as a response to a pressure [94], the density of the medium, temperature and other factors. Fluid media may also form interfaces with solids and other fluids with different density and compressibility. Consequently, the nonlinear response additionally becomes a function of surface and interfacial tension [94].

For example, this is the case of gas bubbles in water that are characterised by a giant value $\beta$ that is several orders of magnitude larger than the values of $\beta$ for water and air [28]. Consequently, whereas in bulk water one can readily observe the generation of five or so higher frequency harmonics of the incident sound [Fig. 2(c)], in water containing gas bubbles one may generate up to 15…20 high-frequency acoustic harmonics by applying the same sound pressure as in the case with bulk water [Fig. 2(d)]. These giant nonlinear properties of gas bubbles have no analogues in optics and they arise from an intrinsic property of bubbles to be easily compressible, which will be discussed in more detail in Section 4.

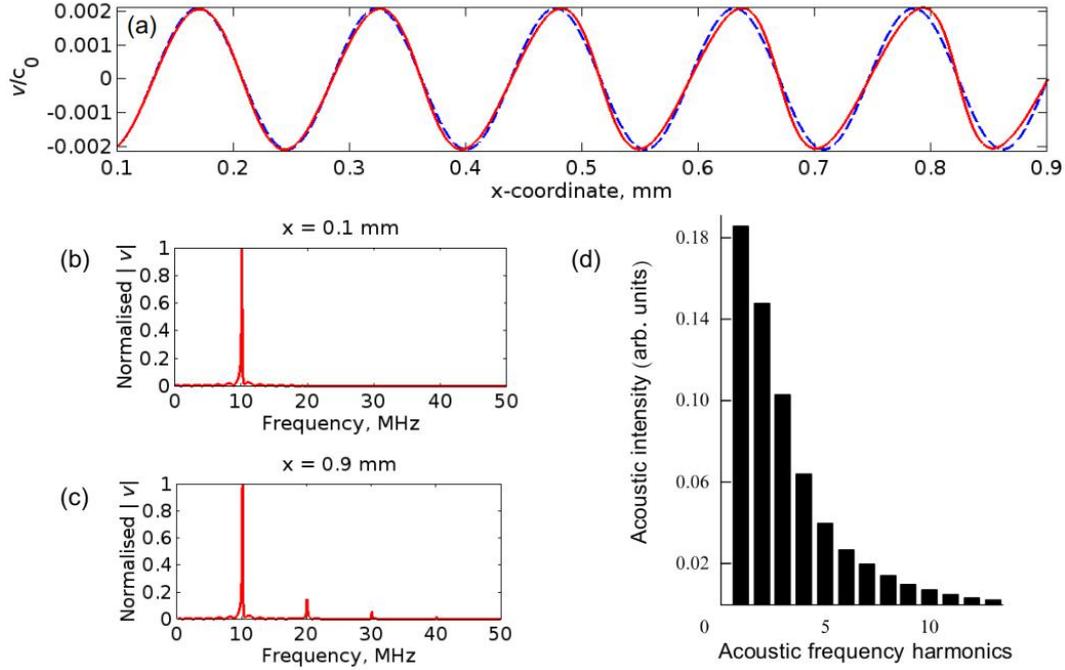

**FIG. 2**: (a) Deformation of the sound wave during its propagation in an acoustically nonlinear medium (here modelled with a finite-difference time-domain method [91] using the nonlinear acoustic coefficient of water β = 3.5). The blue dashed line denotes the same wave but propagating in an idealised (β = 0) medium with the same remaining material parameters. (b-c) Fourier transformation of the acoustic wave performed at two different spatial coordinates. At x = 0.1 mm, the wave is monochromatic with frequency $f_a$ = 10 MHz. New acoustic harmonics appear in the spectrum as the wavefront is deformed. Reprinted from [91]. (d) Giant nonlinear response of a cloud of gas bubbles in a liquid showing the generation of multiple (up to 20 [92]) high-frequency harmonics. Reprinted with permission from [28].

*3.3. Sensing of nonlinear acoustic waves with light*

When a sound wave propagates in a fluid medium, the periodic regions of compression and rarefaction of the particles of the medium lead to changes in the medium density, which also results in changes in the optical refractive index *n* of the medium [95]. Light perceives these changes as a diffraction grating that moves with the speed of sound [95].

This mechanism opens up opportunities to modulate the optical signal with an acoustic wave. If the acoustic wave exhibits nonlinearities, these will also be imprinted onto the modulated optical signal, thereby effectively enabling the conversion of the acoustic nonlinearity into new optical signals [Fig. 1(b)]. For example, this process may reproduce the effect of the nonlinear optical generation of new optical frequencies [Fig. 1(a)].

The physics of the light-sound interaction is in particular important for the development of hydrophones. A hydrophone is a microphone designed to be used underwater for detecting underwater sound. In conventional hydrophones, the acoustic energy mechanically moves a transducer that in turn generates a voltage at the acoustic frequency. Optical hydrophones use frequency modulation of a light beam that Doppler shifts in response to acoustic pressure variations, which also gives rise to the effect of Brillouin light scattering (BLS) from sound [95, 96]. They have an increased sensitivity, greater discrimination against noise and improved directionality as compared with conventional hydrophones [96]. Although the invention of laser played an important role in development of optical hydrophones [96], in practice optical hydrophones may also operate with incoherent sources of light [97], provided the Doppler shift is large enough compared to the linewidth of the source of incoherent light. For example, this condition can be satisfied with gas discharge lamps [95].

Optical hydrophones are also more compact than mechanical ones. The dimensions of state-of-the-art miniaturised optical ultrasonic hydrophones are set by the wavelength of ultrasound in water (~150 μm at 10 MHz) [98-101]. The ability to use incoherent light additionally contributes to development to compact, reliable and inexpensive optical hydrophones.

Optical hydrophones also provide us with a pathway towards using acoustic nonlinearities in optics. However, in this case an optical hydrophone needs to have an effective mechanism for light-sound interaction. We showed that an enhanced strong light-sound interaction can be achieved with a single silver nanorod [91, 102] operating as an antenna for light [78, 103, 104]. (See Section 5 for a more detailed discussion of optical antennas.) The nanorod is immersed into water and insonated by MHz-frequency range ultrasound. In contrast to well-researched GHz-frequency range resonant structural deformations of metal nanostructures such as rods, discs, crosses, and cubes [105-108], in our scenario we investigate the quasi-static regime. This regime of transduction is interesting as some of the dimensions of the nanorod are small compared to the wavelength of light (e.g. the cross-section of the nanorod may be 30 nm × 30 nm and it may be 200…400-nm-long), and all of them are very small compared to the wavelength of MHz-range sound (~150 μm in water). This means that a plasmonic nanoantenna operating as a hydrophone explores a qualitatively different regime from more conventional hydrophone technologies and nano-opto-mechanical systems [105, 106, 107, 108].

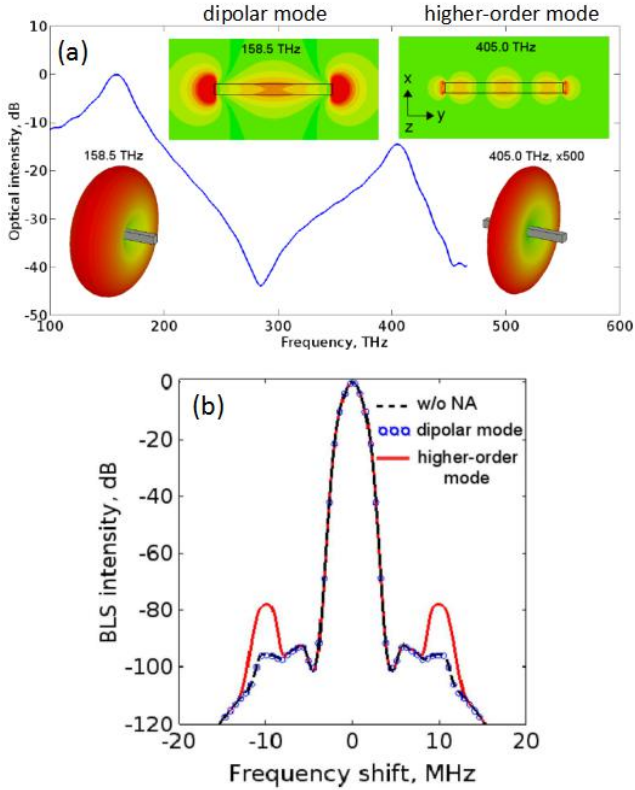

*FIG. 3*: Plasmonic nanoantenna hydrophone. (a) Optical properties of the 340 nm × 30 nm × 30 nm silver nanorod immersed into water. The nanorod plasmonic antenna supports the fundamental dipolar mode and several higher-order modes, one which may be observed at 405 THz. The far-field emission and near-field localisation profiles of light are shown in the insets. Note high emission but poor localisation of light in the fundamental mode regime as compared with the higher-order mode regime. (b) These properties of the higher-order mode lead to much higher sensitivity to MHz-range ultrasound (10 MHz in this case) detected as a frequency shift in the Brillouin Light Scattering (BLS) spectrum. The dashed line marked 'w/o NA' corresponds to the scenario of BLS in bulk water without the nanorod. The linear acoustic regime is assumed in all simulations. Reprinted with permission from [91, 102].

A special design of the nanorod is essential to enable high sensitivity of light to sound, i.e. to changes in the optical refractive index due to sound. The usual operating regime of the nanorod as an optical dipole is unsuitable for sensing of MHz-range ultrasound. This is because in the dipole mode light is localised near the edges of the nanorod [Fig. 3(a)] and hence the interaction of light with sound is small. However, the sensitivity of light to sound is dramatically increased when light bounces back and forth along the nanorod length [Fig. 3(a)]. In this case, the nanorod operates as a nanoscale Fabry-Perot resonator, thereby effectively increasing the light-sound interaction.

These increased sensing properties are confirmed by rigorous simulations of Brillouin light scattering (BLS) [91]. In that particular simulation run, we assume that all acoustic interactions in water are linear. Figure 3(b) shows the simulated BLS spectrum consisting of the central Rayleigh peak and two Brillouin peaks shifted by ±10 MHz, which is the frequency of the incident sound. Three scenarios are considered: (i) the incident light interacts with sound in bulk water and the nanorod is absent, (ii) the nanorod is present but it is tuned on its fundamental optical mode, and (iii) the nanorod operates in the higher-order mode regime. One can see that the intensity of the Brillouin peaks is negligibly small in the first two cases, but the peak intensity, and therefore the sensitivity of light to sound, is much higher when the nanorod is tuned on its higher-order mode. These important observations will be used in the following section to discuss coherent generation of multiple optical frequencies from nonlinear ultrasound.

*3.4. Optical frequency comb generated from nonlinear ultrasound*

Recall that an optical frequency comb is a spectrum consisting of a series of discrete, equally spaced peaks that have a well-defined phase relationship between each other, a phenomenon called phase locking [29-31]. In integrated photonic devices, optical combs are often produced by cascading nonlinear-optical wave mixing processes in a high-Q resonator [12, 63].

In our approach, when optical frequency combs are produced from acoustic nonlinearities, we first obtain equally-spaced acoustic frequency peaks with the same phase, and then convert these peaks into an optical spectrum that inherits the equal frequency spacing and phase-locking [102].

Two physical effects describe the light–sound interaction that underpins the generation of the optical comb: (i) the Bragg reflection of light from the moving diffraction gratings produced by sound waves and (ii) the Doppler shift – change in the frequency of light as the gratings move with respect to the source of light. In a quantum mechanical picture, the interaction of light with sound is an inelastic process where the photon energy is not conserved. As a result, the interaction of light with sound creates Brillouin peaks shifted from the central (Rayleigh) peak by the frequencies of all nonlinear sound waves present in the system [Fig. 4(a)]. The central peak is due to the elastic scattering of light (where photon energy is conserved).

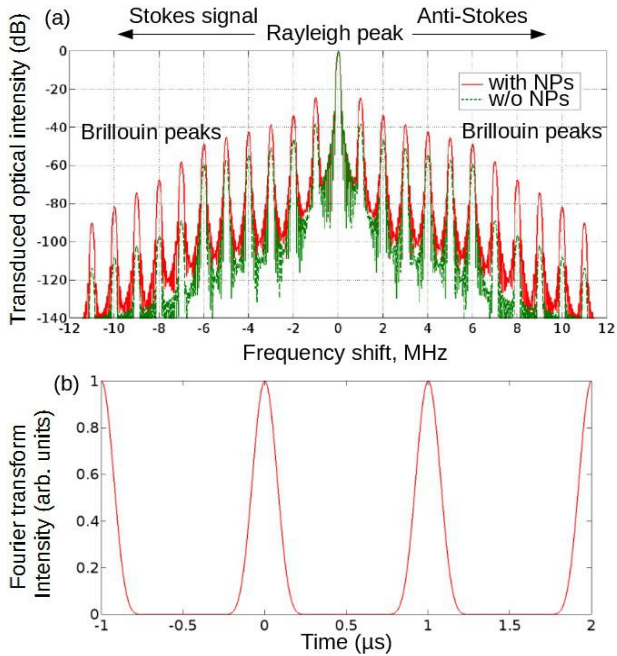

**FIG. 4:** (a) Optical frequency comb generated from nonlinear ultrasound propagating in bulk water (theory). In this particular example, the spacing between the comb peaks equals 1 MHz, which is set by the frequency of the incident sound. Note an enhancement (red solid curve) in the peak intensity due to plasmon nanoparticles (NPs) acting as optical antennas with respect to bulk water without nanoparticles (green dashed curve). (b) Fourier transform of the spectrum in (a). Reprinted with permission from [102].

As shown in Ref. [102], all the peaks in Fig. 4(a) are phase-matched, which is an essential feature of an optical frequency comb [29]. Consequently, similar to mode-locked optical frequency combs, in the time domain the signals at different frequencies should add constructively at one point resulting in a train of optical pulses spaced by $T_p = 1/\delta$ [29], where $\delta$ is the frequency spacing between the peaks in Fig. 4(a). We simulate this physical picture by Fourier-transforming the spectrum in Fig. 4(a) and obtaining time-domain optical intensity signals, which produces well-defined single optical pulses [Fig. 4(b)].

Spectral tuneability of optical frequency combs is required for many real-life applications [11, 31, 32]. Spectral composition of optical frequency combs generated from sound can also be tuned by changing the acoustic frequency. For example, in the particular scenario in Fig. 4(a), the spacing between the comb peaks equals 1 MHz. However, it can be tuned by changing the frequency of the incident sound in a broad spectral range. The lower limit of this range is defined mostly by the linewidth of the optical source. The upper limit is of order of several GHz, which are the frequencies of hypersound [95].

## 4. Giant acoustic nonlinearities of bubbles and droplets

### 4.1. Mechanism of giant acoustic nonlinearities

Liquids have little free space between molecules, which implies that they are not easily compressible. Gases, on the other hand, are easily compressible. Together with the medium's density $\rho$, the compressibility of the fluid medium $\beta_s$ defines the speed of sound in the medium as $c_0 = (\rho\beta_s)^{-1/2}$. We can assume that for water $\rho \approx 1000$ kg/m$^3$ and $\beta_s \approx 4.6 \times 10^{-10}$ Pa$^{-1}$. For air, however, $\rho \approx 1.2$ kg/m$^3$ and $\beta_s \approx 7 \times 10^{-6}$ Pa$^{-1}$, which implies that sound travels faster in water (1474 m/s) than in air (345 m/s).

This distinction also leads to stronger acoustic nonlinearities in systems composed of two fluids with different compressibility and density, and here gas bubbles in water are the most striking example. Let us consider a single gas bubble in otherwise bulk water. When an acoustic wave propagating in bulk water reaches a gas bubble, due to the low density of the gas inside bubble and its high compressibility as compared with water, the amplitude of the acoustic wave is considerably amplified inside the bubble, resulting in dramatic changes in the bubble volume [109]. (Here, we may assume that the bubble maintains its spherical shape [109].) The changes in the bubble's volume result in large acoustic wavefront deformations. As defined above, the deformation of the acoustic wave is quantified by the nonlinear acoustic parameter $\beta$. Therefore, whereas the two components of the bubble − water and air − have $\beta = 3.5$ and $0.7$, respectively, when taken separately [27], an air bubble in water is characterised by $\beta \approx 5000$ [28]. This phenomenon is called giant acoustic nonlinearities [28]. For example, whereas in water one normally observes the generation of about five higher-order acoustic harmonics [Fig. 2(c)], in water with gas bubbles it is possible to observe 15…20 harmonics with the same applied acoustic power and other equal conditions [Fig. 2(d)].

The acoustic response of a single bubble in bulk liquid is described by the nonlinear Rayleigh-Plesset equation [109]. In its simplest form, this equations models a bubble oscillating in an inviscid and incompressible liquid. More complex forms of this equations may take into account the liquid viscoscity and surface tension [109].

For example, Fig. 5(a) shows the solution of Rayleigh-Plesset equation for a 100-nm-radius air bubble in water excited by an ultrasound pulse with a Gaussian-enveloped sinusoid with the frequency $f_a = 50$ MHz. This frequency is detuned from the natural resonance frequency of the bubble $f_0 \approx 30$ MHz, which in the case of water can conveniently be defined as $f_0 R_0 \approx 3$ m/s [109], where $R_0$ is the radius of the bubble at rest.

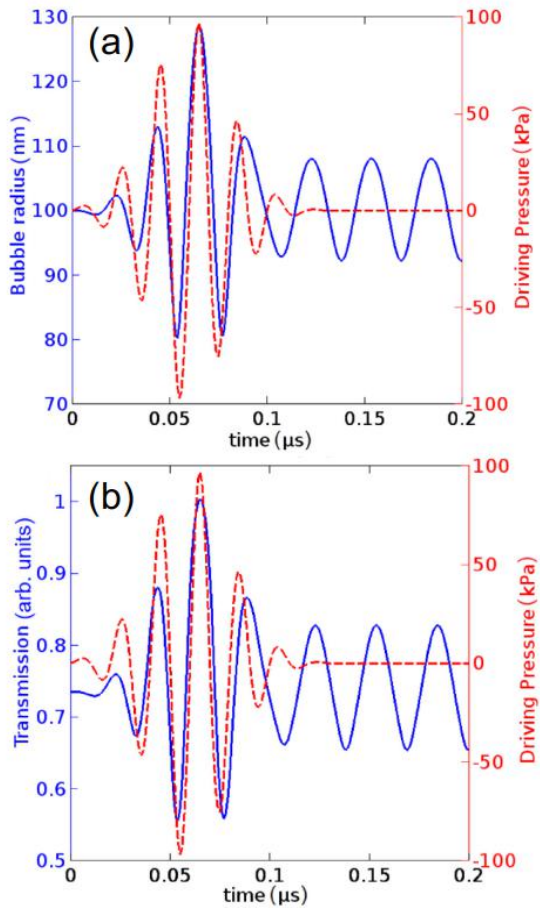

**FIG. 5**: *(a) Theoretical response of a 100-nm-radius air bubble in water (blue solid curve) to the applied acoustic pressure (red dashed curve). (b) The acoustic response of the bubble becomes imprinted onto the intensity of light that illuminates the bubble. Here, the bubble is located in and the light is transmitted through a 300-nm-wide circular, water-filled hole in a 400-nm-thick silver film. Reprinted with permission from [110].*

One can see that the radius of the bubble (blue solid curve) changes in response to the driving pressure (red dashed curve) with a phase lag due to the inertia of the surrounding water [109]. The maximum (minimum) value of the radius reaches ~130 nm (~80 nm). The bubble continues pulsating with a smaller amplitude when the driving pressure signal is turned off. This is because damping is not taken into account in this model. This resonant tail disappears when acoustic losses are taken into account [110].

Significantly, the behaviour of the pulsating bubble modulates the intensity of light scattered by the bubble. This is shown in Fig. 5(b) that plots the calculated intensity of light (blue solid curve) transmitted through a hole in a silver film. The hole is filled with water and it contains a 100-nm-radius air bubble. One can see the shape of the acoustic signal (red dashed curve) imprints onto the light intensity transmitted through the hole.

We now turn our attention to the fact that, in general, the physics of liquid droplets and gas bubbles is similar. Indeed, the surface tension of water results in the wall tension required for the formation of the bubble. The natural tendency to minimise the wall tension forces the bubble to maintain its spherical shape. The same surface tension effect determines the shape of liquid droplets. Because the droplet can be easily deformed, it tends to maintain a spherical shape due to the cohesive forces of the surface layer. As in the case of the bubble, the spherical shape minimises the necessary wall tension.

Thus, liquid droplets oscillate in response to sound similar to gas bubbles and they also exhibit strong nonlinear properties analogous to those of oscillating gas bubbles [111]. However, in the case of droplets, we exploit capillary oscillations that arise because of a competition between inertia of the liquid and surface tension [94, 112]. Unlike acoustic waves that can propagate in all media, capillary oscillations are unique to liquids and they may be driven, for example, electrically or mechanically [113]. Significantly, in contrast to gas bubbles that oscillate and maintain their spherical shape, acoustically-driven liquid droplets often assume complex, non-spherical 3D shapes [Fig. 6(c)] corresponding to specific capillary oscillation modes [112].

It is noteworthy that in droplets we access nonlinear properties because of the softness of the liquid. When a solid-state medium is deformed, the resulting restoring force is due to the stiffness (Young's modulus) of the material. However, in liquid surfaces, the restoring force relies on surface tension [94]. Moreover, the speed of the capillary wave is three orders of magnitude lower than the speed of the acoustic waves in the same liquid [114, 115]. Because of all these distinctions, the frequency of capillary oscillation of a liquid droplet [Fig. 6(a)] is about three orders of magnitude lower than that of the correspondent acoustic mode [Fig. 6(b)] of the same droplet [114, 115].

### 4.2. Interaction of light with droplets and gas bubbles

The optical refractive index of air and water is $n = 1$ and $n = 1.33$, respectively. This refractive index contrast allows detecting single spherical bubbles by light extinction and Mie scattering [109]. A water droplet surrounded by air also has the same refractive index contrast and therefore can be detected using similar techniques. Other optically transparent liquids may be used to form droplets, such as octane ($n = 1.4$), and may form droplets in the form of an emulsion when surrounded by an immiscible liquid.

Thus, when a gas bubble (liquid droplet) interacts simultaneously with light and sound, large changes in the volume of the bubble (or the shape of the droplet) lead to large changes in the behaviour of light [116]. More specifically, the strength of light scattering from sound becomes a function of the complex nonlinear process of compression and rarefaction of the gas inside the bubble (or capillary oscillations of the liquid droplet). This effectively converts nonlinearity into optical signals containing the acoustic frequency components.

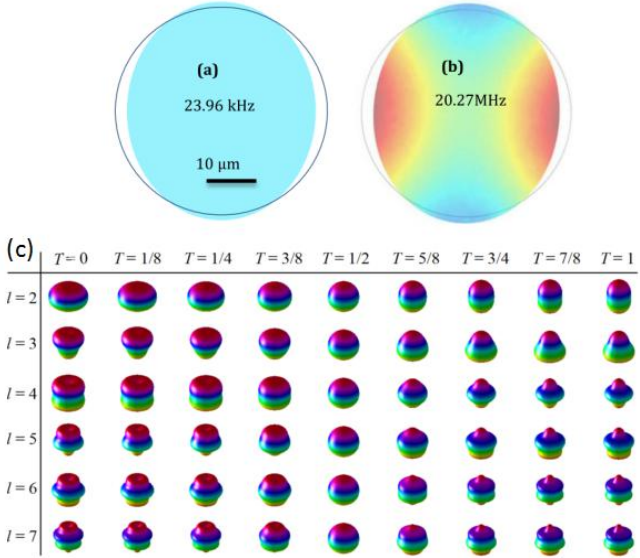

FIG. 6: Comparison between (a) a theoretical capillary oscillation mode of a liquid (silicone oil) sphere in air and (b) its correspondent acoustic mode. The outer circle represents the contour of the sphere at rest and the colour represents pressure. In panel (a), the theoretical shape corresponds to the fundamental capillary oscillation mode (the mode number $l = 2$; $l = 0$ and $l = 1$ are zero-frequency modes corresponding to the conservation of volume and translational invariance, respectively [116]). Reprinted with permission from [114]. (c) Representative oscillation mode shapes of a liquid droplets through a half-period of oscillation for the $l = 2,…,7$ capillary oscillation. $T_l$ is given in units of π radians ($T_l = 1$ is half a period). For the second half of the period the droplet retraces the shapes assumed in the first half.

Although the conversion of acoustic nonlinearity of bubbles and droplets into the optical domain has not been in the focus of previous works, there have been several highly relevant demonstrations of optical signal generation from capillary waves in liquid droplets [117, 118]. The most relevant results demonstrated in those works will be overviewed below.

Before we do that, we note that conceptually similar functionality can also be achieved with well-known acousto-optic modulators that use the acousto-optic effect to diffract and shift the frequency of light using sound waves. For example, in Ref. [119] a nonlinear acousto-optic modulator operating at GHz acoustic (vibrational) frequencies was demonstrated. However, that nonlinearity originates from resonant vibrations of metal nanostructures. This is a different kind of nonlinearity and, although it has important practical applications in its own right [28], it is more difficult to access it with low power due to considerable stiffness of solid-state structures. This is in stark contrast with bubbles and droplets whose softness allows accessing their nonlinear properties with low power.

In Ref. [118] optical tweezers were used to manipulate an octane droplet immersed into water. An optical waveguide placed in close proximity of the droplet was used to deliver the pump light. Radiation pressure from the light circulating in the droplet gives rise to capillary oscillations. Shape changes due to these oscillations result in a Doppler shift of the pump light, thereby giving rise to Stokes emission lines when the pump light is slowly scanned through one of the droplet's optical resonances.

In the resulting frequency spectrum [Fig. 7(a)], the strongest spectral line corresponds to the fundamental mode of the droplet oscillation. (In the notation used in this work, this mode corresponds to the $l = 2$ mode, see the caption to Fig. 6.) The oscillations of the fundamental mode give rise to the second, third and fourth high-frequency harmonics by virtue of nonlinear interactions. In agreement with the theoretical prediction for light scattering from oscillating gas bubbles in water [Fig. 7(b)], the corresponding signal in the time domain is represented by a train of well-defined pulses.

A similar result is observed when the direction of the pump light scan is reversed [Fig. 7(c, d)]. However, in this case the third higher-order capillary oscillation mode of the droplet is excited [$l = 4$ in Fig. 6(c)]. This is because higher-optical-quality-factor modes tend to excite oscillations at a lower rate [120]. The second higher-order mode [$l = 3$ in Fig. 6(c)] is absent because even modes have their node near the equator line [113] and therefore exhibit smaller opto-capillary coupling.

In Ref. [118], the authors proposed to use this effect to build a ripplon laser. The results presented in their paper also open up novel opportunities to generate optical frequency combs from capillary oscillations. It is noteworthy that capillary oscillations may also be excited mechanically (e.g. by ultrasound propagating in a liquid that surrounds the droplet) or electrically (e.g. by using an electrowetting technique [113]).

More generally, the ability to use different excitation mechanisms offers additional degrees of freedom in the design and control of novel multiphysics devices. For instance, considerable attention has recently been paid to exploiting novel regimes of strong coupling between different entities such as photons, phonons and magnons [121, 122]. It has been suggested that multi-resonant photon-phonon-magnon interaction can be achieved with magnetofluidic droplets which combine the softness of fluids with the ability of solids to support resonances of electromagnetic, acoustic and spin waves [123].

Capillary oscillations of liquid droplets are also often detected with laser Doppler vibrometry (LDV). Commercial LDVs often employ a Mach–Zehnder interferometer [Fig. 8(a)] and usually allow characterising the motion of a fluid-fluid interface at frequencies of up to 100 MHz and displacements of as little as a few tens of picometres [124]. Oscillations of the droplet can be excited by either a thickness-mode piezoelectric transducer [PZT, see Fig. 8(b)] or a surface acoustic wave (SAW) device [125].

In Ref. [125], the behaviour of capillary waves in a 2 μL sessile hemispherical water droplet excited by high-frequency acoustic waves oscillating at 500 kHz and 20 MHz was investigated. The droplet was placed on a solid substrate that vibrates to generate capillary oscillation modes [Fig. 8(b)].

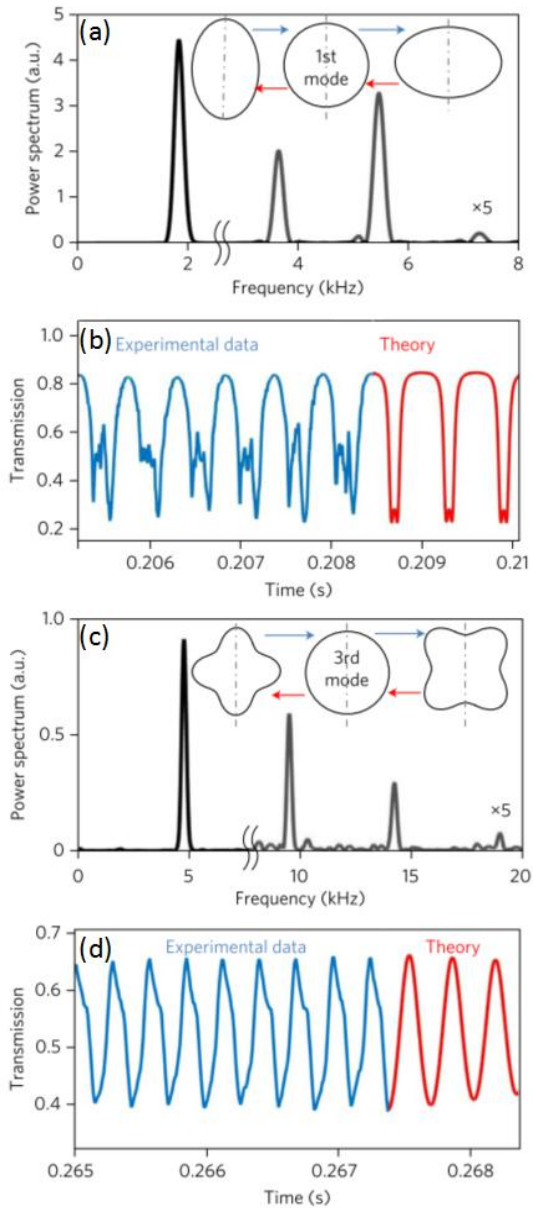

FIG. 7: The interaction of light with different modes of the capillary oscillations of an octane droplet in water leads to the generation of a discrete spectrum with peaks at frequencies of the (a) fundamental and (c) third higher-order oscillation modes. In the time domain (b, d), these spectra are seen as periodic oscillations. Reprinted with permission from [118].

In the measured displacement spectral density [Fig. 8(c)], the resonance peaks at ≥200 Hz are in good agreement with the Lamb model of elastic resonance of a spherical capillary surface [125]. In the 100 Hz − 2 kHz frequency range, the spectrum exhibits a slope predicted by the wave turbulence theory [125]. Significantly, one can also see higher-order resonances from the excitation frequency of 500 kHz upward in a harmonic cascade $f, 2f, 3f,...,7f$, and perhaps beyond; however, these higher-order resonances are not detected due to the upper limit of the LDV's measurement range at that particular resolution

Finally, in this section, we note that oscillations of bubbles [126, 127] and bubble-like objects such as balloons [128] can also be accessed with an LDV.

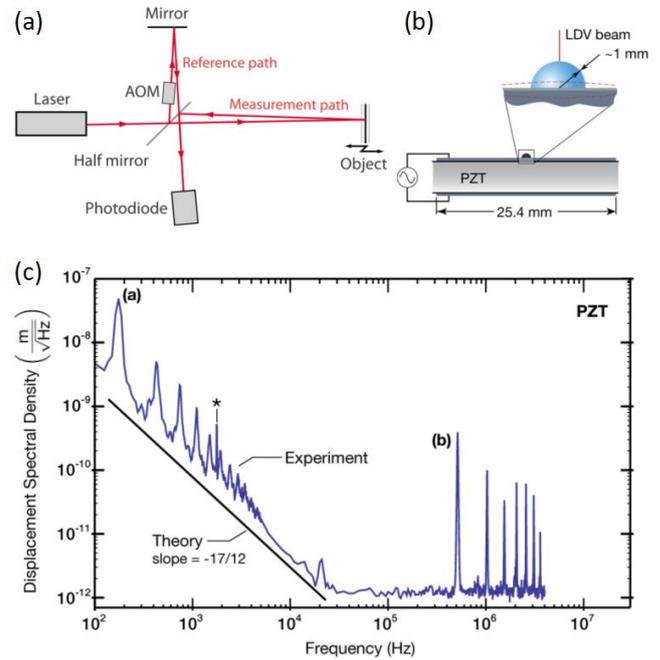

FIG. 8: (a) Schematic of a Mach–Zehnder laser Doppler vibrometer (LDV). The vibrating object at right provides reflection of the laser light. The fixed mirror on the top is a part of the reference path. An acousto-optic modulator (AOM) is used to permit measurement of the objects vibration phase by shifting the reference lasers wavelength by a fixed amount. Reprinted with permission from [124]. (b) Schematic of a thickness-mode PZT. AC power is applied to electrodes on the upper and lower surfaces, inducing a thickness-polarised vibrational mode at 500 kHz [125]. (c) Experimental vibrational spectra for 500 kHz thickness-mode excitation via PZT. Over the frequency range of 100 Hz − 2 kHz, the spectrum exhibits a slope that is close to the value predicted by wave turbulence theory (indicated by the adjacent line). The peak (a) of the spectrum is the fundamental vibrational mode and also the largest response of the capillary wave, though successive resonances may be seen. The peak (b) is the Lorentzian response for the excitation frequency of 500 kHz. Higher-order harmonic resonances at $2f, 3f,..., 7f$ are also observed.

## 5. Plasmon enhancement of light-sound interactions

### 5.1. Plasmonic nanoantennas and nanoparticles

By analogy with a conventional roof-top antenna, an optical nanoantenna emits, receives and, more broadly, controls light with nanoscale elements that are much smaller than the wavelength of the incident light [78, 103, 104]. In particular, nanoscale dimensions and subwavelength operating regime allow using optical nanoantennas to enhance light-matter interaction which, along with other many interesting phenomena useful for sensing, spectroscopy and imaging, leads to strong nonlinear optical effects [42, 68-80]. Optical nanoantennas may be made of metals (most often of gold and silver) or dielectric and semiconductor materials. The former are called plasmonic nanoantennas and the latter are known as dielectric nanoantennas [78].

In plasmonic nanoantennas, enhancement of nonlinear optical phenomena is achieved due to strong light-matter interactions enabled by localised surface plasmon resonances. Plasmon resonances are collective oscillations of the electron charge around metal nanoparticles in resonance with the frequency of light [45]. Localised plasmon resonances lead to trapping of light in close vicinity to nanoparticles, thereby resulting in stronger local optical electric fields and hence stronger nonlinear optical processes (also see Section 2).

The physics of the enhanced light-matter interaction in dielectric nanoantennas is different [78]. Such antennas do not have metal parts and therefore they do not suffer from high light absorption losses and Joule heat production typical of plasmonic nanoantennas. Indeed, plasmons are oscillations of the electron density in metals, and therefore energy is lost and heat is produced when oscillating electrons collide with the metal lattice. In general, the stronger the optical electric field in the metal due to the plasmon resonance, the stronger the losses and heat.

However, light localisation and enhancement in dielectric nanoantennas may be several orders of magnitude smaller than in their plasmonic counterparts. Nevertheless, the achievable enhancement levels can enable strong nonlinear optical effects, especially because light is localised inside the body of the nanoantenna [129].

The interaction of light with sound in liquids may be increased when gold or silver nanoparticles operating as optical nanoantennas are present in the liquid [91]. Returning to the example of the optical frequency comb generation from ultrasound propagating in bulk water in the nonlinear regime. As shown in Fig. 4(a), the presence of a single silver nanorod particle in water gives rise to a ~100-fold increase in the intensity of the peaks in the generated BLS spectrum. An ensemble of such particles acting as an array of plasmonic nanoantennas will further increase the magnitude of the peaks. For example, this may be achieved due to high light intensity 'hot spots' in the gaps between the nanoparticles [78].

As demonstrated above, the interaction of light with acoustically-driven gas bubbles in water leads to generation of new optical frequencies. The nanoantenna-assisted increase in the sensitivity of light to sound will also work in this scenario. Indeed, microscopic gas bubbles in liquids can be integrated with metal nanoparticles that are located in the shell of the bubble, thereby leading to the effect of a 'metal bubble' [130, 131]. Such gas bubbles have already been employed as a bimodal biomedical imaging agent that provides contrast for both ultrasonic and optical imaging. In particular, to be useful for optical imaging, the optical resonance of 'metal bubbles' falls within the visible-to-near-infrared spectral range [131].

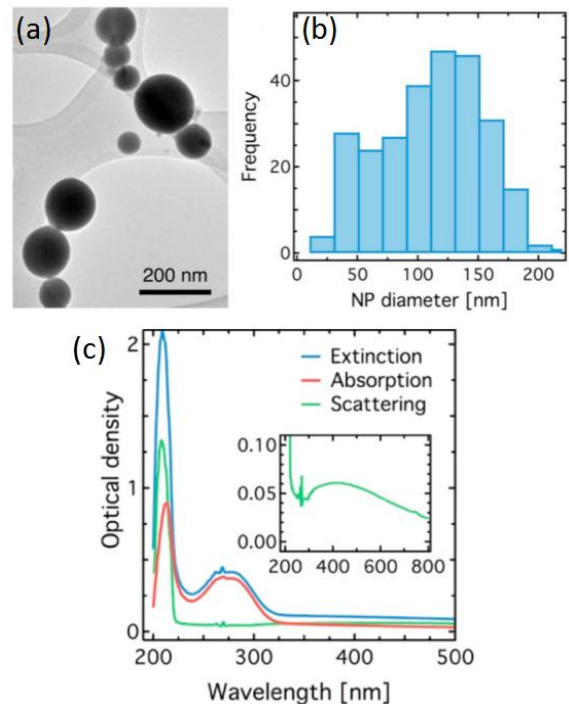

*FIG. 9*: *(a) TEM image and (b) size distribution of EGaIn liquid-metal nanoparticles suspended in ethanol. (c) Experimental extinction, absorption and scattering spectra showing the plasmon resonances at ~215 nm and ~270 nm. Inset: Scattering spectrum in a wider wavelength range. Reprinted from [150].*

*5.2. UV plasmonic liquid-metal nanoparticles*

Thus far, we have discussed the opportunities to use plasmonic resonances to increase the interaction of light with giant acoustic nonlinearities of gas bubbles in liquids. However, it would also be beneficial to combine the plasmonic properties with the capillary oscillations of liquid droplets. Unfortunately, however, water and other common liquids cannot support surface plasmon resonances.

To circumvent this problem, one may employ room-temperature liquid metals such as gallium and its alloys [132-134]. Such metals are excellent candidates for the role of plasmonics nanodroplets (also called liquid-metal nanoparticles).

Nanoparticles made of non-noble metals such as gallium have recently attracted significant interest due to promising plasmonic applications in the UV-visible spectral range [135-145]. Unlike gold and silver, gallium has a Drude-like dielectric permittivity function extending from the UV range through the visible and, mostly in the liquid state, into the infrared spectral region [140, 146]. Due to their high environmental stability and excellent mechanical properties gallium and its alloys are highly relevant for many emerging applications in electronics, micro-mechanics and chemistry [134, 147, 148].

Significant effort has recently been done to synthetise colloidal liquid-metal nanoparticles [133, 149]. Consequently, we investigated eutectic gallium-indium (EGaIn, 75% Ga 25% In by weight) nanoparticles suspended in ethanol [Fig. 9(a, b)] and demonstrated their strong plasmonic resonances in the UV spectral region [Fig. 9(c)] [150]. EGaIn has negligible vapour pressure and low viscosity [147, 148], which is important for excitation of high-intensity capillary oscillation modes. Consequently it allows strongly modulating the UV plasmon spectra of liquid-metal nanoparticles with ultrasound [Fig. 10(a)].

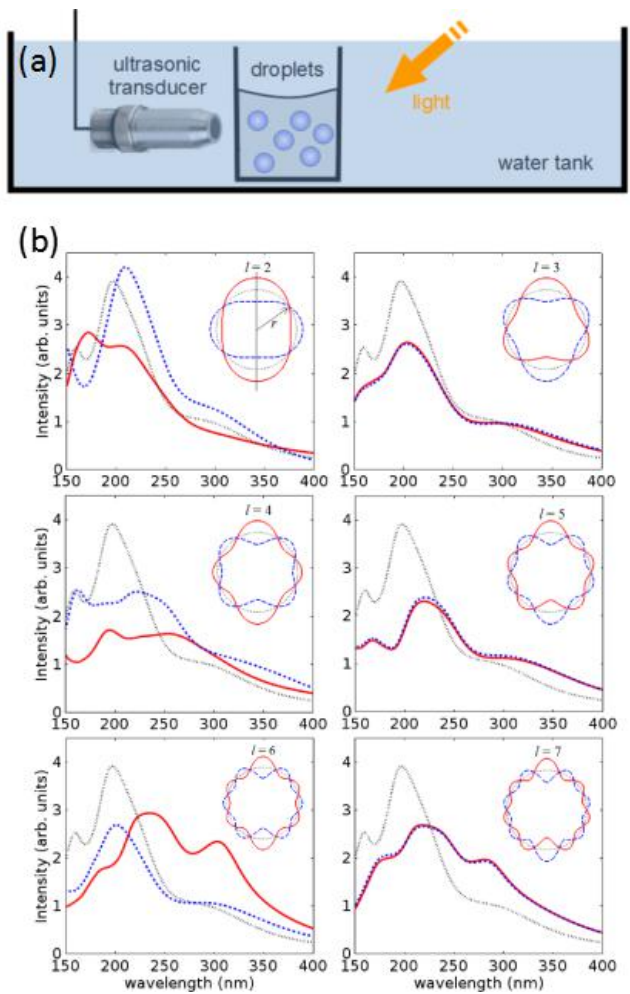

*FIG 10: (a) Schematic of an experimental setup for the excitation of capillary oscillation of colloidal liquid-metal nanoparticle with ultrasound. (b) Scattered light intensity spectra calculated for static 3D shapes assumed by the nanoparticle at T = 0 (dashed curves), T = 1 (solid curves), and T = 1/2 (dotted curve) shown in Fig. 6(b). The insets show the cross-sections of these 3D shapes. The same curve styles are used in the insets and the main panels. Reprinted with permission from [116].*

Figure 10(b) shows the calculated light scattering spectra of the liquid-metal nanoparticles tuned on one of the attainable capillary oscillation modes. One can see a dramatic change in the UV plasmonic spectrum of the nanoparticles that oscillates as a function of the driving acoustic pressure. These changes enable the conversion of nonlinear acoustically-driven oscillations into optical signals in the UV range.

## 6. Regimes of extreme nonlinear acoustic events

As demonstrated in Section 4.2, radial oscillations of gas bubbles and effective acoustic response of water containing multiple gas bubbles are highly nonlinear. More broadly, a bubble in a liquid can be considered as a nonlinear dynamical system exhibiting complex resonances, multistability, bifurcations to chaos and other complex processes [109]. For example, by driving a 10 μm radius-at-rest gas bubble with a considerable sound pressure amplitude of 300 kPa and a high frequency of 600 kHz, one may achieve chaotic radial oscillations represented as a complicated set with a fractal structure known as a strange attractor [109]. The oscillations of a gas bubble can also switch to a new oscillation regime with twice the period of the original oscillation system, a process called period doubling [109]. Moreover, with an even stronger acoustic driving pressure, the response of the bubble may develop hysteresis.

Consequently, the conversion of these complex nonlinear phenomena into the optical domain may result in new intriguing fundamental physics that would complement the purely optical, complex nonlinear phenomena such as optical bistability [151-153] and low-power optical frequency comb generation exploiting optical period-doubling bifurcation [87].

Another intriguing possibility associated with the giant nonlinearity of gas bubbles is to exploit extreme processes that accompany the effect of bubble cavitation [109]. Cavitation is the formation of vapour bubbles in a liquid due to forces acting on the liquid. Very often the cavitation occurs when a liquid is subjected to rapid changes of pressure that cause the formation of bubbles where the pressure is relatively low. When subjected to higher pressure, the bubbles implode and can generate an intense shock wave. Experimental results also demonstrate that collapsing bubbles reach temperatures of ~5000 K, pressures of ~1000 atm, and heating and cooling rates above $10^{10}$ K/s [154]. These events can create extreme physical and chemical conditions in otherwise cold liquids, such as ice breaking [155].

Cavitation also occurs, and can be controlled, in the presence of an acoustic pressure wave. When microscopic gas bubbles are present in the liquid in which the acoustic wave propagates, the bubbles are forced to oscillate due to the applied acoustic pressure. With sufficiently high acoustic pressures, the bubbles will first grow in size and then rapidly collapse [156, 157].

A gas bubble collapsing near an interface exhibits especially interesting properties [158-160]. In general, a bubble collapsing near a solid interface develops a jet directed towards the interface [blue curves in Fig. 11(a)]. This jet is so strong that it can lead to cavitation damage of the solid surface [159, 160]. In the case of a free surface and fluid-fluid interfaces [158], the jet is directed mostly away from the surface but the free surface or the fluid-fluid interface undergo considerable deformations due to the evolution of the bubble [Fig. 11(b)]. The collapse of a bubble near a free surface or a fluid-fluid surface may also result in the formation of new bubbles [158]. The results in Fig. 11 were produced by using computational code that simulates asymmetric bubble cavitation [161]. The fluid is assumed to be incompressible, inviscid and irrotational, and surface tension is neglected. The Navier-Stokes equations are solved using the boundary integral method [162].

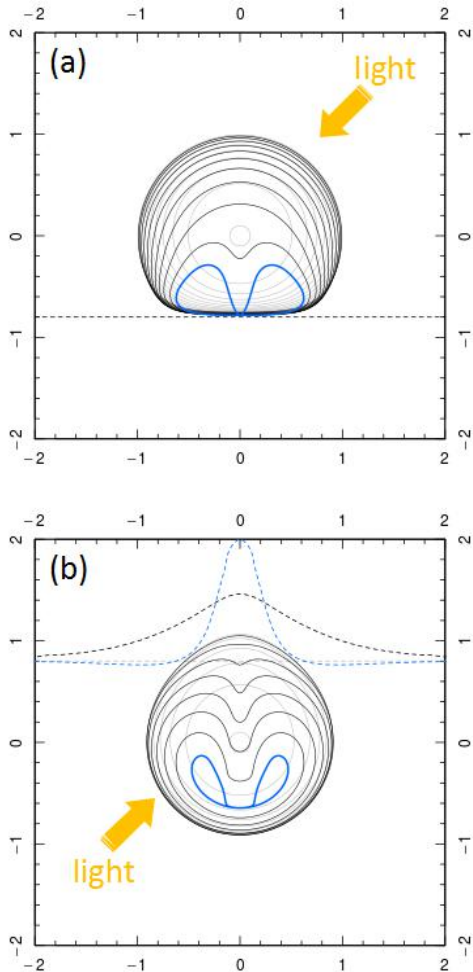

*FIG. 11*: *Simulation of an asymmetric cavitation bubble during its growth (thin grey curves), collapse (thick black curves) and jet-impact (blue curves). (a) The bubble is located near a rigid surface (dashed line). (b) The bubble is located near a free surface in (dashed curves). Note that the free surface is perturbed by the changes in the bubble's shape and its eventual collapse. The coordinates are in the units of the maximum bubble radius [161]. The interaction of the incident light with the bubble and the adjacent surface provide access to giant nonlinearities. Simulations were carried out with a computational code developed by Danail Obreschkow, the University of Western Australia [161].*

The evolution of the bubble near an interface is highly nonlinear [163, 164]. This implies that the displacement of both the surface of the bubble and free surface are a nonlinear function of the driving acoustic pressure. Light can readily sense these displacements. Furthermore, an even higher sensitivity may be achieved in a system where a gas microbubble collapses near a liquid-metal surface. Here, the liquid-metal is expected to behave as a liquid that responds the oscillations and collapse of the bubble similar to the scenario in Fig. 11(b). However, at the same time the liquid-metal interface may support surface plasmon modes [145], which will allow for efficient sensing of small, potentially nanometre-range, fluctuations of the metal surface.

**Conclusions and outlook**

Modern nonlinear photonics is mostly impossible without lasers. In this review paper, we have shown how a judiciously engineered interaction between sound, capillarity and low-power incoherent light provides access to exciting and yet largely unexploited nonlinear phenomena of non-optical origin. Understanding, controlling and harnessing these new nonlinear phenomena promises to remove the restriction of high-power laser light operation. This also enables compact and readily deployable photonic devices for telecommunications, sensing and biomedical applications.

For example, although lasers have important applications in biomedicine (e.g. in imaging, skincare, and kidney stone removal [165]), classical nonlinear optical processes have had relatively little impact in the field of biomedicine [166] because high laser powers required to achieve optical nonlinearites cause photodamage to living cells and tissues [13, 14]. Hence, effectively nonlinear, low-power photonics devices would usher a new class of biologically friendly photonic sensors and imaging systems.

We have also demonstrated that acoustic nonlinearity can be accessed with gas bubbles and liquid droplets that oscillate when subjected to acoustic waves. Significantly, gas bubbles have already found important applications in biomedicine and were approved by the Food and Drug Administration (FDA) and similar organisations as ultrasound contrast agents operating inside the human body [167, 168]. These dramatic progress will provide a considerable advantage for novel biomedical optical devices and imaging modalities exploiting the giant nonlinear acoustic properties of bubbles.

Of course, the use of fluids as a constituent material for photonic devices presents a number of technological challenges, such as the integration of liquid-state elements into traditional solid-state microphotonic systems [169, 170]. Nevertheless, liquid-state devices can offer a number of potentially transformative advantages for microphotonic systems [114, 115, 118]. Moreover, there have been successful examples of liquid-state and gas-state devices that are compatible with conventional solid-state photonics such as liquid-state optical lenses [171], liquid-core optical fibres used in spectroscopy [172] and gas-filled optical fibres used to generate optical frequency combs [173, 174]. In our Ref. [102], we further capitalise on the progress in the liquid-core optical fibre development and suggest to employ liquid-core fibre filled with water containing gas micro-bubbles.

Further technological challenges may include the need to align solid metal nanoparticles inside a liquid to optimise the excitation of plasmon resonance modes. This may be required when plasmonic effects are exploited to increase the strength of the light-sound interaction. In solid-state plasmonic devices, the position and also the orientation of nanoparticles with respect to the polarisation of the incident light are fixed. In a liquid-state devices, however, external disturbances and temperature changes may result in movement and changes in the nanoparticle orientation. To circumvent these effects, one may employ magneto-plasmonic nanoparticles made of ferromagnetic metals [175-177]. By applying an external static magnetic field to the resulting magneto-fluid, one may control not only the positioning of the nanoparticles but also their collective optical response, thereby resulting in intriguing magneto-optical effects [178, 179] and important applications in biomedical imaging (see Ref. [177] for a relevant discussion).

Finally, the proposed application of acoustic nonlinearities in photonics further contributes to the ongoing effort to complement and strengthen optical nonlinearities with nonlinear effects of other nature such as, for example, nonlinearities of spin waves [180]. These developments may lead to interesting fundamental physics because both sound and spin wave excitations can simultaneously interact with light [123].

**Acknowledgements**
We thank Antony Orth, Brant Gibson, Philipp Reineck, Leslie Yeo, Lillian Lee, Amgad Rezk (RMIT University), Mark Hutchinson (University of Adelaide), Michael Dickey (North Carolina State University, USA), Bradley Boyd (University of Canterbury), and Dr Richard Nowak (Olympus Australia) for valuable discussions. This work was supported by the Australian Research Council (ARC) through its Centre of Excellence for Nanoscale BioPhotonics (CE140100003), LIEF program (LE160100051), and Future Fellowship (FT160100357, FT180100343). This research was undertaken with the assistance of resources and services from the National Computational Infrastructure (NCI), which is supported by the Australian Government.